\documentclass{article}
\usepackage{graphicx}
\usepackage[centertags]{amsmath}
\usepackage{amsfonts}
\usepackage{amssymb}
\usepackage{amsthm}
\title{Symmetries and constraints in mechanics of continua}

\author{Jan J. S\l awianowski\\
Institute of Fundamental Technological Research,\\
Polish Academy of Sciences,\\
$5{\rm B}$, Pawi\'{n}skiego str., 02-106 Warsaw, Poland\\
e-mail: jslawian@ippt.gov.pl}

\begin{document}

\maketitle
\begin{abstract}
In spite of its long history and classical character which goes back even to d'Alembert and Lagrange, the problems of constraints in mechanics of continua is still mysterious and full of misunderstandings. Let us mention the problem of difference between special solutions of unconstrained systems and constraints dynamics, holonomic and nonholonomic constrained, various versions of so-called vakonomy, sub-Riemannian structures, servoconstraints, and programme motion. There are some strange confusions between field theory and mechanics of continua; many problems call for conceptual cleaning, although both the mechanics of continua and field theory are formally based on partial differential equations. The author is deeply indebted to professor Czes\l aw Wo\'{z}niak for his help during many years, and deep understanding and support 
\end{abstract}

\section{Constrained dynamics versus constrained solutions}
  
The problem of constraints in mechanics is very delicate and full of confusions and misconceptions. Situation is maximally clear in mechanics of system with a finite number of degrees of freedom, subject to holonomic constraints. Let us consider a system of $N$ material points moving in $n$-dimensional Euclidean space $M$. "Physically" $n=3$, but let us forget about this. If some other symbols turn out to be needed, the $N$-dimensional linear space of translations in $M$ will be denoted by $V$, and the metric tensor by $g\in V^{*}\otimes V^{*}$; obviously, $g$ is symmetric and positively definite.

Let the force acting on the $A$-th material point be denoted by 
\[
\overline{F}_{A}(\overline{r}_{1}, \ldots, \overline{r}_{N};\overline{v}_{1},\newline  \ldots, \overline{v}_{N};t) 
\]
the explicit dependence on time $t$ is admissible. Obviously, $\overline{r}_{A}$ is the radius-vector of the $A$-th material point (with respect to some fixed origin in $M$) and $\overline{v}_{A}=d\overline{r}_{A}/dt$ is its velocity vector (independent on the origin, but evidently dependent on the choice of the inertial reference frame). Unconstrained equations of motion have the obvious Newton form:
\begin{equation}\label{EQ1}
m_{A}\frac{d^{2}\overline{r}_{A}}{dt^{2}}=\overline{F}_{A}\left(\overline{r}_{1}, \ldots, \overline{r}_{N};\frac{d \overline{r}_{1}}{dt}, \ldots, \frac{d \overline{r}_{N}}{dt};t\right).
\end{equation}
Kinetic energy is given by
\begin{equation}\label{EQ2}
T=\frac{1}{2}\sum_{A}m_{A}\frac{d \overline{r}_{A}}{dt}\cdot\frac{d \overline{r}_{A}}{dt}
\end{equation}
with the obvious symbol for the scalar product:
\begin{equation}\label{EQ3}
\overline{u}\cdotp\overline{v}=g(\overline{u},\overline{v})=g_{ij}u^{i}w^{j}.
\end{equation}

If we use generalized coordinates $Q^{a}$ and the corresponding kinematical metric tensor $G$ on the configuration space $Q:=M^{N}$,
\begin{equation}\label{EQ4}
G_{ab}=\sum_{A}m_{A}\frac{\partial \overline{r}_{A}}{\partial Q^{a}}\cdot\frac{\partial \overline{r}_{A}}{\partial Q^{b}},
\end{equation}
and its contravariant inverse $G^{ab}$,
\begin{equation}\label{EQ5}
G^{ac}G_{cb}=\delta^{a}{}_{b},
\end{equation}
equations of motion become
\begin{equation}\label{EQ6}
\frac{D^{2}Q^{a}}{Dt^{2}}=F^{a}\left( \ldots, Q^{b}, \dots; \ldots \frac{dQ^{b}}{dt}; t\right).
\end{equation}
Here generalized accelerations are given by 
\begin{equation}\label{EQ7}
\frac{D^{2}Q^{a}}{Dt^{2}}=\frac{d^{2}Q^{a}}{dt^{2}}+\Gamma^{a}{}_{bc}\frac{dQ^{b}}{dt}\frac{dQ^{c}}{dt},
\end{equation}
$\Gamma^{a}{}_{bc}$ are Christoffel coefficients for $G$,
\begin{equation}\label{EQ8}
\Gamma^{a}{}_{bc}=\frac{1}{2}G^{ai}\left( G_{ib,c}+G_{ic,b}-G_{bc,i}\right),
\end{equation}
comma denotes the partial differentiation with respect to $Q^{a}$, and generalized forces are given by 
\begin{equation}\label{EQ9}
F^{a}=\sum_{A}G^{ab}\overline{F}_{A}\cdotp\frac{\partial \overline{r}_{A}}{\partial q^{b}}.
\end{equation}
Obviously, the configuration kinematical metric $G$ is here flat, i.e., its curvature tensor does vanish. Nevertheless, equations (\ref{EQ6}), (\ref{EQ7}) are more general and hold also for the general Riemann space $(Q, G)$, used as a configuration space.

For the potential system, the forces $\overline{F}_{A}$ are given by
\begin{equation}\label{EQ10}
F^{i}{}_{A}=-g^{ij}\frac{\partial V}{\partial r^{j}{}_{A}}=-g^{ij}\frac{\partial V}{\partial r^{j}{}_{B}}\delta_{BA}.
\end{equation}

In terms of generalized coordinates:
\begin{equation}\label{EQ11}
F^{a}=-G^{ab}\frac{\partial V}{\partial Q^{b}}.
\end{equation}
Equations of motion are then derivable from the variational principle:
\begin{equation}\label{EQ12}
\delta I=\delta \int L dt=0,
\end{equation}
where $L=T-V$, and have the Euler-Lagrange form
\begin{equation}\label{EQ13}
\frac{\delta I}{\delta Q^{a}}=\frac{\partial L}{\partial Q^{a}}-\frac{D}{Dt}\frac{\partial L}{\partial \dot{Q}^{a}}=0;
\end{equation}
the symbol $D/Dt$ denoting the total (substantial) derivative with respect to time $t$; do not confuse it with the covariant differentiation. 

If in addition to the Lagrangian background some other forces $\phi^{a}$, e.g., dissipative ones (friction) are present, then (\ref{EQ13}) is replaced by
\begin{equation}\label{EQ14}
\frac{\partial L}{\partial Q^{a}}-\frac{D}{Dt}\frac{\partial L}{\partial \dot{Q}^{a}}=G_{ab}\phi^{b}.
\end{equation}

The problem of constraints is as follows: In addition to explicitly given forces like $\overline{F}_{A}$, $F^{a}$, etc., there are additional ones, which confine motion to some submanifold $W\subset Q$ of dimension $f< \dim Q$. More precisely, this is the problem of holonomic constraints. The manifold $W$ may be analytically described by the system of equations:
\begin{equation}\label{EQ15}
F_{a}(\ldots, Q^{b}, \ldots)=0, \quad a=1, \ldots, m=\dim Q-f.
\end{equation}

And now some very important point comes, namely one concerning the distinction between two different problems:
\bigskip

{\bf Procedure 1}. Problem of special solutions of (\ref{EQ13})/(\ref{EQ14}) satisfying equations (\ref{EQ15}), i.e., placed on
\begin{equation}\label{EQ16}
W:=\{z\in Q: F_{a}(z)=0, a=1, \ldots, m\}
\end{equation}
(it is implicitly assumed that $\phi_{a}$ are functionally independent, at least in a neighbourhood of $W$).

{\bf Procedure 2}. Problem (\ref{EQ13})/(\ref{EQ14}) dynamically modified by (\ref{EQ15}). This problem consists in that in addition to a priori given forces (\ref{EQ1}), (\ref{EQ9}), (\ref{EQ10}), etc., there are some additional ones, usually even not described explicitly (usually, but not always), $\overline{\mathcal{R}}_{A}$, $\mathcal{R}_{a}$ which result in small-amplitude (but sometimes large-energy and high-frequency) oscillations vertical to $W$. Their orthogonal projections onto $W$ do not satisfy original equations, but something else. 
\bigskip

Obviously, it is well-known that the total, joint system consisting of the primary equations of motion in any of the versions (\ref{EQ1}), (\ref{EQ13}), (\ref{EQ14}) and constraining conditions (\ref{EQ15}) is inconsistent: as a rule, it is intrinsically contradictory. There are only exceptional situations when inconsistency does not appear. For example, when the forces $\overline{F}_{A}$ do vanish, i.e., motion is free, and $W$ is an affine submanifold of the Euclidean space $Q$ (functions $F_{a}$ are affine, i.e., roughly speaking, linear-inhomogeneous), then all straight-lines swept with constant velocities are solutions (they form the general solution of the problem of motion in $W$). But of course, the situation changes drastically when the general constraints of rigid motion are imposed onto the free motion problem. Then the only solutions of the above {\bf Procedure 1} are the total free translations. But we are aware that with a good approximation the general rigid motion with nontrivial rotations does exist and is both theoretically interesting and practically relevant; one knows that from elementary school. The point is that one deals then with the above {\bf Procedure 2}. The solution is classical, known from ages and belongs mainly to Lagrange and d'Alembert. Nevertheless, there are situations when this classical procedure is forgotten. From some point of view this is the problem of passive control, based on the natural contact-rolling-friction mechanism, without servomechanism-aided and computer-aided procedures and active control elements, cf \cite{1}--\cite{6}. From some point of view the
idea is peculiar. Namely, instead of real auxiliary forces orthogonal
to $W$, defined around $W$, and just vanishing at $W$ itself, one
introduced symbolic d'Alembert reactions which are just defined on
$W$ itself and maintain the among-constraints motion. Therefore,
(\ref{EQ1}) is replaced by
\begin{equation}
m_{A}\frac{d^{2}\overline{r}_{A}}{dt^{2}}=\overline{F}_{A}
\left(\overline{r}_{1},\ldots,\overline{r}_{N};\frac{d\overline{r}_{1}}{dt},
\ldots\frac{d\overline{r}_{N}}{dt};t\right)+\overline{\mathcal{R}}_{A}
\left(\overline{r}_{1},\ldots,\overline{r}_{N};\frac{d\overline{r}_{1}}{dt},
\ldots\frac{d\overline{r}_{N}}{dt};t\right)\label{EQ17}
\end{equation}
and this equation is completed and treated jointly with constraint
equations (\ref{EQ15}) 
\begin{equation}
F_{a}\left(\overline{r}_{1},\ldots,\overline{r}_{N}\right)=F_{a}
\left(\ldots,Q^{b},\ldots\right)=0\label{EQ18}
\end{equation}
But of course the joint system (\ref{EQ17}) \& (\ref{EQ18}) is over-determined
unless some constitutive conditions are imposed on $\overline{\mathcal{R}}_{A}$.
The d'Alembert-Lagrange procedure of passive control tells us that
the along-constraints motion is neither affected nor maintained without
the energy transfer to the system, i.e., reactions $\overline{\mathcal{R}}_{A}$
do not do any work, i.e., their mechanical power does vanish; they
are orthogonal to the constraints surface $W$, i.e., to all virtual
motions:
\begin{equation}
\mathcal{P}_{\mathcal{R}}=\underset{A}{\sum}\overline{\mathcal{R}}_{A}
\cdot\frac{d\overline{r}_{A}}{dt}=0\label{EQ19}
\end{equation}
for any functions $\mathbb{R}\ni t\rightarrow\overline{r}_{A}\left(t\right)$
satisfying equations:
\begin{equation}
F_{a}\left(\overline{r}_{1}\left(t\right),\ldots,\overline{r}_{N}\left(t\right)\right)=0\qquad\textrm{identically over}\quad t,\label{EQ20}
\end{equation}
therefore, for any virtual velocities $\frac{d\overline{r}_{A}}{dt}$
subject to
\begin{equation}
\frac{\partial F_{a}}{\partial\overline{r}_{A}}\cdot\frac{d\overline{r}_{A}}{dt}=0.\label{EQ21}
\end{equation}
Obviously, the symbols $\partial F_{a}/\partial\overline{r}_{A}$
are abbreviations for the systems of derivatives with respect to the
components $r^{i}{}_{A}$. This means that reactions $\overline{\mathcal{R}}_{A}$
are given by:
\begin{equation}
\overline{\mathcal{R}}_{A}=\underset{a}{\sum}\lambda^{a}\frac{\partial F_{a}}{\partial\overline{r}_{A}}\label{EQ22}
\end{equation}
with the same meaning of $\partial F_{a}/\partial\overline{r}_{A}$
as above. Here $\lambda_{a}$ are some apriori unknown Lagrange multipliers.
The total system of equations of motion in the sense of the {\bf Procedure 2} above, consists of (\ref{EQ17}), (\ref{EQ20}), (\ref{EQ22}). This is the system of $f+m=\dim Q+m$ equations
for the system of $\dim Q+m$ variables $\overline{r}_{A}$, $\lambda^{a}$.
Eliminating $\lambda^{a}-s$ we obtain the system of $f$ independent
equations for $f$ independent generalized coordinates parametrizing
$W\subset Q$. The knowledge of $\lambda^{a}$, i.e., reactions $\overline{\mathcal{R}}_{A}$ is essential for the design of the system of factors maintaining the constraints.

Let us go back to a more homogeneous notation. We consider analytical
mechanics in a differential manifold $Q$ of dimension $\dim Q=k$,
with the Lagrangian background $L:\: TQ\rightarrow\mathbb{R}$ ( $TQ$
denoting the tangent bundle of $Q$, i.e., manifold of generalized
coordinates $q^{i}$ and velocities $v^{i}=\dot{q}^{i}$), and with
certain non-Lagrangian forces, e.g., dissipative ones, represented
in the covariant terms as $D_{i}$. If $Q$ is endowed with Riemannian
structure $G$, and something like the ``magnetic field'' $A_{i}$
(covector potential) is present, and the scalar potential $V$ as
well, then the typical school example is:
\begin{equation}
L\left(q,v\right)=\frac{1}{2}G_{ij}\left(q\right)v^{i}v^{j}+\varepsilon A_{i}\left(q\right)v^{i}+V\left(q\right).\label{EQ23}
\end{equation}
Typical model of dissipative forces is 
\begin{equation}
D_{i}\left(q,v\right)=-d_{ij}\left(q,v\right)v^{j},\label{EQ24}
\end{equation}
where $d_{ij}\left(q,0\right)=0$, and $d_{ij}$ is symmetric and
positively definite. In applications it often does not depend on coordinates
$q^{i}$.

Quite independently on the above particular structure, in general
equations of motion of the system $\left(L,D\right)$ have the form:
\begin{equation}
\frac{D}{Dt}\frac{\partial L}{\partial\dot{q}^{i}}-\frac{\partial L}{\partial q^{i}}=D_{i},\qquad i=1,\ldots,\dim Q=k.\label{EQ25}
\end{equation}

Let us quote, although it is not particularly important, that for
(\ref{EQ23}) the resulting field equations have the form: 
\begin{equation}
\frac{D^{2}q^{i}}{Dt^{2}}=\varepsilon G^{ik}F_{kj}\frac{dq^{j}}{dt}-G^{ik}\frac{\partial V}{\partial q^{k}}-G^{ik}d_{kj}\frac{dq^{j}}{dt}\:,\quad i=1,\ldots, k,\label{EQ26}
\end{equation}
where 
\begin{equation}
F_{kj}=A_{j,k}-A_{k,j}=\frac{\partial A_{j}}{\partial q^{k}}-\frac{\partial A_{k}}{\partial q^{j}}.\label{EQ27}
\end{equation}
Holonomic constraints are described analytically by functionally independent
equations 
\[
F_{a}\left(q^{1},\ldots,q^{k}\right)=0\:,\quad a=1,\ldots, m,
\]
and the total system of equations of motion is given by
\begin{eqnarray}
\frac{D}{Dt}\frac{\partial L}{\partial\dot{q}^{i}}-\frac{\partial L}{\partial q^{i}}=D_{i}+\mathcal{R}_{i}, &  & F_{a}\left(q^{1},\ldots,q^{k}\right)=0,\label{EQ28}\\
i=1,\ldots, k, &  & a=1,\ldots, m,\nonumber 
\end{eqnarray}
together with the condition that reactions $\mathcal{R}_{i}$ are
passive, i.e., do not any work on virtual motions compatible with
constraints, i.e., such ones that
\begin{equation}
\underset{i}{\sum}\mathcal{R}_{i}\left(q\right)\frac{dq^{j}}{dt}=0\label{EQ29}
\end{equation}
if equations 
\begin{equation}
F_{a}\left(q^{1}\left(t\right),\ldots,q^{k}\left(t\right)\right)=0\:,\quad a=1,\ldots, m,\label{EQ30}
\end{equation}
are satisfied. Let us notice that $q^{i}(t)$ in (\ref{EQ30}) are quite
arbitrary (excepting, of course, appropriate smoothness conditions).
This implies that reactions are given by:
\begin{equation}\label{EQ31}
\mathcal{R}(q, v)_{i}=\sum_{a}\lambda^{a}(q, v)\frac{\partial F_{a}}{\partial q^{i}}, \quad i=1, \ldots, k.
\end{equation}
The total system of equations
\begin{equation}\label{EQ32}
\frac{D}{Dt}\frac{\partial L}{\partial \dot{q}^{i}}-\frac{\partial L}{\partial q^{i}}=D_{i}+\sum_{a}\lambda^{a}\frac{\partial F_{a}}{\partial q^{i}}, \quad i=1, \ldots, k,
\end{equation}
\[
F_{a}(q^{1}, \ldots, k^{k})=0, \quad a=1, \ldots, m,
\]
consists of $(k+m)$ independent conditions imposed on $(k+m)$ functions of time, $q^{i}(t)$, $\lambda^{a}(t)$. And, roughly speaking, $\lambda^{a}$ are to be eliminated, and the general solution for the time dependence of $q^{i}$ is to be determined. Although, more precisely, in engineering design the quantities $\lambda^{a}$ are very relevant in the analysis of the endurance of mechanical factors responsible for constraints (various "rods", "threads", etc.).

This is implicit description. In certain problems the explicit parametric formulation is more convenient. So, let the constraints $W\subset Q$ be parametrically given by:
\begin{equation}\label{EQ33}
q^{i}=\varphi^{i}(y^{1}, \ldots, y^{k-m}), \quad i=1, \ldots, k;
\end{equation}
the parameters $y^{\mu}$, $\mu=1, \ldots, f=k-m$ are independent "proper" coordinates on $W\subset Q$, i.e., they label "true" degrees of freedom of the constrained system. Then our equations of motion may be written as follows, in explicitly irredundant terms:
\begin{equation}\label{EQ34}
\frac{D}{Dt}\frac{\partial L_{W}}{\partial \dot{y}^{\mu}}-\frac{\partial L_{W}}{\partial y^{\mu}}=D_{W\mu}, \quad \mu=1, \ldots, f=k-m,
\end{equation}
where, obviously, $L_{W}$ is the restriction of the original $L$ to the tangent subbundle $TW$, and $D_{W}$ is the pull-back of $D$ to $W\subset Q$,
\begin{equation}\label{EQ35}
L_{W}(y, \dot{y}):=L\left( \varphi^{i}(y), \frac{\partial \varphi^{i}}{\partial z^{\nu}}\dot{z}^{\nu}\right),
\end{equation}
\[
D_{W\mu}(y, \dot{y}):=D_{i}\left( \varphi^{j}(y), \frac{\partial \varphi^{j}}{\partial z^{\nu}}\dot{z}^{\nu}\right)\frac{\partial \varphi_{i}}{\partial z^{\mu}}.
\]

\section{Nonholonomy and vakonomy}

It is interesting to mention here briefly about non-holonomic constraints, although it is not our main subject here \cite{1}--\cite{10}.

Non-holonomic constraints based on the natural slide-free rolling on rough surfaces are linear in generalized velocities; in certain situations one deals with "linear non-homogeneous", i.e., affine conditions,
\begin{equation}\label{EQ36}
\mathcal{F}_{a}(q, v)=\omega_{ai}(q)v^{i}+f_{a}(q)=0, \quad a=1, \ldots, m.
\end{equation}
Of course, holonomic constraints (\ref{EQ30}) may be also formally written in this way, then $\omega_{ai}$ are given by derivatives of $F_{a}$-s in (\ref{EQ32}):
\begin{equation}\label{EQ37}
\omega_{ai}=\frac{\partial F_{a}}{\partial q^{i}}, \quad a=1, \ldots, m.
\end{equation}

There are also mixed situations, when both the motion in $Q$ and instantaneous virtual velocities are independently restricted, and one deals with the mixture of genuine non-holonomic, and holonomic constraints. Let us do not go here into such details, roughly, the true non-holonomy is assumed here to follow from the
non-integrability of the Pfaff problem:
\begin{equation}
\omega_{a}=\omega_{ai}\left(q\right)dq^{i}=0\:,\quad a=1,\ldots,m.\label{EQ38}
\end{equation}
This means that at any configuration $q\in Q$ we are given some linear
$(k-m)$-dimensional subspace $W_{q}\in T_{q}M$ of virtual velocities.
There are various degrees of non/integrability. There are two extreme
situations: quasiholonomic constraints, when $Q$ is foliated (stratified)
by the $m$-dimensional family of $(k-m)$-dimensional strata (fibers),
and the total non-holonomy, when at least locally, at any $q\in Q$
there is a neighbourhood $U\subset Q$ such that any of its points
may be approached from $q$ along a curve compatible with (\ref{EQ36})/(\ref{EQ38}) (compare this with Caratheodory formulation of the Second Principle of Thermodynamics).

According to the d'Alembert principle, the reactions maintaining such
con\-straints are ideal, i.e., energetically passive, they do not do
any work along virtual motions compatible with constraints. The total
system of equations of motion is given by 
\begin{equation}
\frac{D}{Dt}\frac{\partial L}{\partial\dot{q}^{i}}-\frac{\partial L}{\partial q^{i}}=D_{i}+\lambda^{a}\omega_{ai}\:,\quad\omega_{ai}\frac{dq^{i}}{dt}=0;\label{EQ39}
\end{equation}
obviously, the summation convention meant with respect to repeated
indices.

This is again the system of $(k+m)$ equations for $(k+m)$ variables $(q^{i}, \lambda^{a})$. But in non-holonomic theory some new problems appear which some more than one century ago resulted in a big confusion, misunderstanding, simply shame. Namely, in the case of variational system, the Lusternik theorem for the variational problem
\begin{equation}\label{EQ40}
\delta\int L(q(t), \dot{q}(t))dt=0
\end{equation}
constrained by the conditions
\begin{equation}\label{EQ41}
F_{a}(q(t), \dot{q}(t))=\omega_{ai}(q(t))\dot{q}^{i}(t)=0
\end{equation}
results in equations:
\begin{equation}\label{EQ42}
\frac{D}{Dt}\frac{\partial L}{\partial \dot{q}^{i}}-\frac{\partial L}{\partial q^{i}}=R^{i}, \quad \omega_{ai}\frac{dq^{i}}{dt}=0,
\end{equation}
where, however, the reactions are not any longer given by the right-hand sides of (\ref{EQ39}). 

Lusternik theorem, i.e., confined extremum (confined stationary point) after some easy calculations implies that
\begin{equation}\label{EQ43}
R_{i}=\mu^{a}\left( \frac{\partial \omega_{aj}}{\partial q^{i}}- \frac{\partial \omega_{ai}}{\partial q^{j}}\right)\frac{dq^{j}}{dt}-\frac{d\mu^{a}}{dt}\omega_{ai}.
\end{equation}

This is something evidently different than the dissipative-free (\ref{EQ39}). The obvious generalization of (\ref{EQ42}) to the dissipative case is
\begin{equation}\label{EQ44}
\frac{D}{Dt}\frac{\partial L}{\partial\dot{q}^{i}}-\frac{\partial L}{\partial q^{i}}=D_{i}+R_{i}.
\end{equation}
Only the second term in (\ref{EQ43}) is analogous to (\ref{EQ39}), when we identify the holonomic Lagrange multiplies $\lambda^{a}$ with $d\mu^{a}/dt$. The system (\ref{EQ43}) \& (\ref{EQ44}) consists of $(k+m)$ differential equations imposed on $(k+m)$ variables $(q^{i}, \mu^{a})$. Let us notice however, that this is essentially a system of differential equations, because the Lagrange multipliers enter there in a differential, non-algebraic way. In this sense the system is more "elastic". And equations of motion  (\ref{EQ43})/(\ref{EQ44}) are evidently different than (\ref{EQ32}). Moreover (\ref{EQ43})/(\ref{EQ44})  do not describe rough, slide-free rolling of natural mechanical systems. The "magnetic"-like term controlled by $\mu$ is completely strange from the point of view of such applications. The system based on  (\ref{EQ43})/(\ref{EQ44}) has evidently more degrees of freedom. Such system were called by  Russian school "vakonomic" (variational axiomatic). In Western literature they are called "sub-Riemannian". Quite unexpected applications were found in financial mathematics. Besides, such systems are interesting from the point of view of pure differential geometry. It seems that systems of this kind may be used in active control, especially when one deals with servomechanisms and computer-aided problems of programme motion. Systems based on variational principles and higher-order Lusternik principles (higher-order differential, but also integral and functional), have some special features interesting from the point of view of energetic balance in active control. In particular, this is the case with non-holonomic constraints nonlinear in velocities, accelerations and higher-order time derivatives. Certainly they are non-physical in natural mechanism of sliding-free constraints, but they are promising from the point of view of active control. 

In automatic and active control some first-order constraints nonlinear in velocities may be used, e.g., when stabilizing velocity of satellites and space ships.

If such constraints are given by equations:
\begin{equation}
F_{a}(q,v)=0, \qquad a=1, \ldots, m, \label{EQ45}
\end{equation}
then there are good geometric reasons to control the system with the Appell-Chetajev reactions:
\begin{eqnarray}
\frac{D}{Dt}\frac{\partial L}{\partial \dot{q}^{i}} - \frac{\partial L}{\partial q^{i}} = D_{i}+ \lambda^{a} \frac{\partial F_{a}}{\partial v^{i}}, \qquad F_{a}(q,v)=0, \\ \nonumber
a=1, \ldots, m, \qquad i=1, \ldots, k. \label{EQ46}
\end{eqnarray}

In any case, the Appell-Chetajew reactions are geometrically correctly-de\-fined (in a manner independent on the choice of coordinates). 

There are also physical reasons to expect the physical utility from the variational Lusternik procedure. The corresponding equations of motion have the form:
\begin{equation}
\frac{D}{Dt}\frac{\partial L}{\partial \dot{q}^{i}} - \frac{\partial L}{\partial q^{i}} = D_{i}+R_{i} \qquad F_{a}(q,\dot{q})=0, \label{EQ47}
\end{equation}
with the program forces / reactions given by:
\begin{equation}
R_{i}= \mu^{a}\frac{\partial F}{\partial q^{i}} - \frac{d \mu^{a}}{dt}\frac{\partial F_{a}}{\partial \dot{q}^{i}} - \mu^{a} \frac{\partial^{2} F_{a}}{\partial \dot{q}^{i} \partial q^{j}} \frac{d q^{j}}{dt}- \mu^{a} \frac{\partial^{2} F_{a}}{\partial \dot{q}^{i} \partial \dot{q}^{j}} \frac{d^{2} q^{j}}{dt^{2}} =0. \label{EQ48}
\end{equation}

They are also geometrically correctly defined and it is clear that the main term has just the Appell-Chetajew form $\frac{d \mu^{a}}{dt}\frac{\partial F_{a}}{\partial \dot{q}^{i}}$. The last term in (\ref{EQ48}) represents the control of inertial properties. 

\section{Constraints and symmetries in deformable \\ bodies}

What is not clear with constraints if everything is so clear and classical? It seems that people doing with continua often do not distinguish between {\bf Procedures 1} and {\bf 2} in Section 1 above. The point is that mechanics of continua is often confused with field theory. In spite of using partial differential equations in both disciplines, they are something else. There is only one obvious exception. When dealing with incompressible (isochoric) media, one usually does not forget about the pressure Lagrange multiplier which is just the reaction force responsible for incompressibility. But one forgets about the problem in other models, and it is a great merit of professor Czes\l aw Wo\'{z}niak \cite{11}--\cite{14} that he stressed the problem. Incidentally, the author here is very indebted to professor Wo\'{z}niak for his understanding the problem, and his permanent support. First of all, let us notice that from the geometric point of view, the most interesting and important constraints are ones implied by some symmetry groups. 

These are usually some groups responsible for geometry of the physical space or space time, like isometry group, affine group, conformal group, Poincare group, Galilei group, etc. Configuration space of various constrained continua very often happen to be homogeneous spaces of those groups. One of examples which was very interesting for us was affinely rigid body, i.e., body rigid in the sense of affine geometry, homogeneously deformable body. It was the model of internal degrees of freedom in Eringen's micromorphic continuum. There are also other interesting examples, like molecular vibrations, collective models of nuclei, astrophysical objects, geophysical problems, macroscopic elasticity in situations when the wave length is comparable with the linear size of the body etc. And here there is plenty of misunderstandings. Namely, one often does not distinguish between {\bf Procedures 1}, {\bf 2} from Section 1. One is often faced with the statement that there is only a small family of solutions. The point is that one confuses {\bf Procedure 1} with {\bf Procedure 2} and one looks for the special solutions of unconstrained problems, rather then on the constrained dynamics with its characteristic reaction forces. An extremely strange argument is that both the deformation tensor and the stress tensor are constant within the homogeneously deformable body, because of which motion is to be trivial.

Obviously, finite bodies with boundary cannot have constant deformation tensor, except its interior. And the reaction analysis shows that, in virtue of the d'Alembert principle, reactions responsible for the affine rigidity do not vanish, however, it is their monopole and dipole distributions that vanishes, i.e., therefore the total reaction force and the dipole distribution of reactions do vanish. Because of this, if the configuration of affine body is given by
\begin{equation}
x^{i}(r,\varphi; t)= r^{i}(t)+ \varphi^{i}{}_{K}(t)a^{K}, \label{EQ49}
\end{equation}
where $r^{i}$ are coordinates of the centre of mass, $\varphi^{i}{}_{K}$ are internal/relative parameters, and $a^{K}$ are material variables, then equations of motion have the form:
\begin{equation}
M \frac{d^{2}r^{i}}{dt^{2}}=F^{i}, \qquad \varphi^{i}{}_{K} \frac{d^{2}\varphi^{j}{}_{L}}{dt^{2}}J^{KL}= N^{ij}. \label{EQ50}
\end{equation}
The meaning of symbols is as follows:
\begin{itemize}
	\item $M$ is the total mass of the body,
\begin{equation}
M=\int d\mu \label{EQ51}
\end{equation}
	\item $J^{KL}$ is the co-moving tensor of inertia in the material space, thus constant, 
\begin{equation}
J^{KL}= \int a^{K}a^{L}d\mu(a) \label{EQ52}
\end{equation}
	\item center of mass is placed at $a^{K}=0$,
\begin{equation}
J^{K}= \int a^{K}d\mu(a)=0 \label{EQ53}
\end{equation}
	\item $F^{i}$ is the total force,
\begin{equation}
F^{i}= \int \mathcal{F}^{i}(a)d\mu(a) \label{EQ54}
\end{equation}
	\item $N^{KL}$ is the co-moving dipole of forces distribution, therefore, its spatial/Eule\-rian components are given by
\begin{equation}
N^{ij}= \int \varphi^{i}{}_{K}\varphi^{j}{}_{L}a^{K}a^{L}d\mu(a)=  \varphi^{i}{}_{K}\varphi^{j}{}_{L} \int a^{K}a^{L}d\mu(a). \label{EQ55}
\end{equation}
\end{itemize}

Let us quote some alternative forms of equations of motion as balance laws for linear momentum $p$ and affine spin $K$
\begin{equation}
\frac{dp^{i}}{dt}=F^{i} \qquad \frac{dK^{ij}}{dt}= \frac{d \varphi^{i}{}_{K}}{dt} \frac{d \varphi^{j}{}_{L}}{dt}J^{KL} + N^{ij}, \label{EQ56}
\end{equation}
where
\begin{equation}
p^{i}= M\frac{dr^{i}}{dt}, \qquad K^{ij}= \varphi^{i}{}_{K}\frac{d \varphi^{j}{}_{L}}{dt} J^{KL} \label{EQ57}
\end{equation}
are respectively translational momentum and affine spin. In other words:
\begin{equation}
\frac{dp^{i}}{dt}=F^{i}, \qquad \frac{dK^{ij}}{dt}= \Omega^{i}{}_{m}K^{mj}, \label{EQ58}
\end{equation}
\begin{equation}
\Omega^{i}{}_{j}=\frac{d\varphi^{i}{}_{A}}{dt}\varphi^{-1A}{}_{j}, \label{EQ59}
\end{equation}
is an affine velocity, i.e., Eringen's "gyration"
\begin{equation}
\widehat{\Omega}^{A}{}_{B}= \varphi^{-1Ai}{}_{i}\varphi^{j}{}_{B}\Omega^{i}{}_{j}, \label{EQ60}
\end{equation}
is its co-moving representation.

Let us also quote the following formula:
\begin{equation}
\frac{dK^{ij}}{dt}=N^{ij}+2 \frac{\partial T_{int}}{\partial g_{ij}}, \label{EQ61}
\end{equation}
where the kinetic energy is given by
\begin{equation}
T=T_{tr}+T_{int}= \frac{M}{2}g_{ij} \frac{dr^{i}}{dt} \frac{dr^{j}}{dt}+ \frac{1}{2}g_{ij} \frac{d\varphi^{i}{}_{K}}{dt} \frac{d\varphi^{j}{}_{L}}{dt}J^{KL}. \label{EQ62}
\end{equation}

If Lagrangian is given by
\begin{equation}
L=T-V\left(r^{i}, \varphi^{i}{}_{K} \right), \label{EQ63}
\end{equation}
then 
\begin{equation}
p_{i}, K^{i}{}_{j} \label{EQ64}
\end{equation}
are respectively Hamiltonian generators of spatial translations and affine rotations about the centre of mass.
\begin{equation}
S^{ij}=K^{ij}-K^{ji}\label{EQ65}
\end{equation}
is the spin angular momentum, and 
\begin{equation}
\frac{dS^{ij}}{dt}=N^{ij}-N^{ji},\label{EQ66}
\end{equation}
thus, spin is conserved if $N^{ij}$ is symmetric

Let us quote a few additional interesting formulas: 
\begin{equation}
\frac{d\widehat{p}^{A}}{dt}=-\widehat{p}^{B}J_{BC}K^{CA}+\widehat{F}^{A},\qquad \frac{d\widehat{K}^{AB}}{dt}=-\widehat{K}^{AC}J_{CD}\widehat{K}^{DB}+N^{AB},
\label{EQ67}
\end{equation}
where symbol with the capital indices denote co-moving component of
physical quantities and 
\begin{equation}
J_{AC}J^{CB}=\delta_{A}{}^{B}.\label{EQ68}
\end{equation}
Another geometrically interesting expressions: 
\begin{equation}
M\frac{d\widehat{v}^{A}}{dt}=M\widehat{\Omega}^{A}{}_{B}\widehat{v}^{B}
+\widehat{F}^{A},\qquad\frac{d\widehat{\Omega}^{B}{}_{C}}{dt}J^{CA}=
-\widehat{\Omega}^{B}{}_{D}\widehat{\Omega}^{D}{}_{C}J^{CA}
+\widehat{N}^{AB}.\label{EQ69}
\end{equation}

It is clear that we deal here with the system of $n^{2}+n=n(n+1)$
degrees of freedom (in the $n$-dimensional space; physically it is
12, when $n=3$) and this is just the dimensionality of the general
solution, according to the {\bf Procedure 2} in Section 1.

There are another interesting problems concerning dynamical affine
invariance in mechanics of affinely rigid bodies and nonholonomic
constraints, both usual and VAKONOMIC in mechanics of affine bodies,
however, there is no place for them here. Some of them are discussed
in \cite{2}, \cite{3}, \cite{6}. 

Let us notice that if the forces are potential, then the dynamical
quantities (\ref{EQ56}) (\ref{EQ57}) are given by
\begin{equation}
F^{i}=-g^{ij}\frac{\partial V}{\partial r^{j}}\qquad N^{ij}=-\varphi^{i}{}_{A}\frac{\partial V}{\partial\varphi^{k}{}_{A}}g^{kj}.\label{EQ70}
\end{equation}

If there exist dissipative forces non-derivable from Lagrangian or
Hamiltonian, then in addition to (\ref{EQ56})some additional terms appears.
In the simplest case, we choose them linear or quadratic in generalized
velocities $dr^{i}/dt$, $d\varphi^{i}{}_{k}/dt$.

Let us observe another interesting point, namely, some additional,
geometric, i.e., group-implied forces imposed onto (\ref{EQ56}). Gyroscopic
constraints, or rather pseudo-holonomic constraints of rigid motion,
consist of the first equation of (\ref{EQ56}), i.e., the assumption that
$\Omega^{i}{}_{j}$, $\widehat{\Omega}^{A}{}_{B}$ are respectively
$g$-skew-symmetric and $\eta$-skew-symmetric angular velocities in spatial and co-moving representations, 
\begin{equation}
\Omega^{i}{}_{j}=-\Omega_{j}{}^{i}=-g_{ik}\Omega^{k}{}_{i}g^{ij},\qquad \widehat{\Omega}^{A}{}_{B}=-\widehat{\Omega}_{B}{}^{A}=
-\eta_{BC}\widehat{\Omega}^{C}{}_{D}\eta^{DA},\label{EQ71}
\end{equation}
where $g$ is the metric tensor of the physical space and $\eta$
is the material (reference) metric. The conditions (\ref{EQ56}) are then
evidently holonomic and may be written down as the conditions of isometry,
\begin{equation}
g_{ij}\varphi^{i}{}_{A}\varphi^{j}{}_{B}=\eta_{AB}.\label{EQ72}
\end{equation}

This is explicitly purely holonomic form. Then the reaction moments
$N_{R}$ are evidently symmetric,
\begin{equation}
\left.N_{R}\right._{ij}=\left.N_{R}\right._{ji},\label{EQ73}
\end{equation}
and equations (\ref{EQ56}) are evidently free of explicitly non-specified reactions. Gyroscopic reactions do not vanish, however their full tensor contractions with skew-symmetric affine virtual velocities (angular velocities) are vanishing in virtue of constraints (\ref{EQ71}), (\ref{EQ72}). Taking skew-symmetric part of (\ref{EQ56}) we eliminate reaction moments and obtain the effective equations of motion.

Let us now consider isochoric constraints, i.e., incompressibility. Here one is faced with something traditionally very familiar and important in continuum mechanics, first of all in fluids. The traces of affine velocities do vanish then:
\begin{equation}\label{EQ74}
{\rm Tr}\ \Omega=\Omega^{i}{}_{i}=0.
\end{equation}

The total contractions of such virtual $\Omega$-s with the reaction affine moment $N_{R}$ must vanish:
\begin{equation}\label{EQ75}
N_{R}{}^{ij}\Omega_{ji}=N_{R}{}^{ij}\Omega^{k}{}_{i}g_{jk}=0.
\end{equation}
But this means that reactions are pure traces,
\begin{equation}\label{EQ76}
N_{R}{}^{i}{}_{j}=\lambda\delta^{i}{}_{j}, \quad N_{R}{}^{ij}=\lambda g^{ij}, \quad \lambda=\frac{1}{n}{\rm Tr}. N_{R}=\frac{1}{n}g_{ij}N_{R}{}^{ij}.
\end{equation}

Therefore, to eliminate the Lagrange multiplier $\lambda$, we must take the constraints condition (\ref{EQ74}) (i.e., $\det \varphi=$const) jointly with the $g$-traceless part of (\ref{EQ56}) itself, i.e., explicit,
\begin{equation}\label{EQ77}
\varphi^{i}{}_{A}\frac{d^{2}\varphi^{j}{}_{B}}{dt^{2}}J^{AB}-\frac{1}{n}g_{ab}\varphi^{a}{}_{A}\frac{d^{2}
\varphi^{b}{}_{B}}{dt^{2}}J^{AB}g^{ij}=N^{ij}-\frac{1}{n}g_{ab}N^{ab}g^{ij}.
\end{equation}

One can discuss constraints implied by the linear-conformal group, i.e., generated by rotations and dilatations. Then affine velocity (gyration) has the form:
\begin{equation}\label{EQ78}
\Omega^{i}{}_{j}=\omega^{i}{}_{j}+\alpha\delta^{i}{}_{j},
\end{equation}
where $\omega^{i}{}_{j}$ is the $g$-skew-symmetric angular velocity, i.e., it satisfies (\ref{EQ71}) and $\alpha$ is an arbitrary real, dilatational parameter, so that
\begin{equation}\label{EQ79a}
g_{ij}\varphi^{i}{}_{A}\varphi^{j}{}_{B}=\lambda\eta_{AB}, \quad \lambda>0.
\end{equation} 
Then reaction-free equations of motion consist of the skew-symmetric part of (\ref{EQ56}) and of the $g$-trace of that equation, so
\begin{equation}\label{EQ80}
\varphi^{i}{}_{A}\frac{d^{2}\varphi^{j}{}_{B}}{dt^{2}}J^{AB}-\varphi^{j}{}_{A}\frac{d^{2}\varphi^{i}{}_{B}}{dt^{2}}J^{AB}=N^{ij}-N^{ji},
\end{equation}
\begin{equation}\label{EQ80a}
g_{ij}\varphi^{i}{}_{A}\frac{d^{2}\varphi^{j}{}_{B}}{dt^{2}}J^{AB}=g_{ij}N^{ij}.
\end{equation}

Reaction moments $N_{R}{}^{ij}$ are symmetric and $g$-traceless.

And finally some very interesting example of non-holonomic, but non-VA\-KONOMIC constraints, when $\Omega$ is $g$-symmetric. Those are constraints of the purely rotation-free motion (the only geometrically correct definition);
\begin{equation}\label{EQ81}
\Omega^{i}{}_{j}-\Omega_{j}{}^{i}=\Omega^{i}{}_{j}-g_{jk}g^{il}\Omega^{k}{}_{l}=0.
\end{equation} 
Reactions are anti-symmetric, and (\ref{EQ81}) must be joined with the symmetric part of (\ref{EQ56}),
\begin{equation}\label{EQ82}
\varphi^{i}{}_{A}\frac{d^{2}\varphi^{j}{}_{B}}{dt^{2}}J^{AB}+\varphi^{j}{}_{A}\frac{d^{2}\varphi^{i}{}_{B}}{dt^{2}}J^{AB}=N^{ij}+N^{ji}.
\end{equation}

Think the motion of suspension in a viscous fluid as an example. There are also VAKONOMIC models of this kind. But this a quite different story, no place for it here.

\section*{Acknowledgements}

As mentioned, I am very indebted to my older Friend Professor Wo\'{z}niak, for everything he cordially did for me during my years of scientific work. 

It is also a happy coincidence; for some reasons I just wanted to write some work about constraints, and the opportunity of the anniversary of Professor Wo\'{z}niak turned out to be very happy for me in this respect.

During many years, in all my KBN and Ministry of Education grants, I permanently returned to ides by Professor Wo\'{z}niak and his cordial support.

\end{document}